Corresponding author:   S. Robaszkiewicz, Institute of Physics A. Mickiewicz University, Umultowska 85, 61-614 POZNAŃ, POLAND, tel. +0048-61-8257-492, fax. +0048-61-8257-758, e-mail: saro@main.amu.edu.pl

# Effects of diagonal disorder on Charge Density Wave and Superconductivity in local pair systems

Grzegorz Pawłowski and Stanisław Robaszkiewicz

Institute of Physics, A. Mickiewicz University, Poznań, Poland

## Abstract

We analyse the influence of diagonal disorder (random site energy) on Charge Density Wave (CDW) and Superconductivity (SS) in local pair systems which are described by the model of hard core charged bosons on a lattice. This problem was previously studied within the mean field approximation for the case of half filled band ($n = 1$). Here we extend that investigation to the case of arbitrary particle concentration ($0 < n < 2$) and examine the phase diagrams of the model and the behaviour of superfluid density ($\rho_s$) as a function of n and the increasing disorder. Depending on the strength of random on-site energies, the intersite density-density repulsion and the concentration the model can exhibit several various phases, including homogeneous phases: CDW, SS and Bose-glass (NO) as well as the phase separated states: CDW-SS, CDW-NO and particle droplets. The obtained results for SS phase are in qualitative agreement with the available Monte Carlo calculations for two dimensional lattice.

Also, in a definite range of parameters the system exhibits the phenomena which we call a disorder induced superconductivity and a disorder induced charge ordering.

## 1. Introduction

The purpose of the present work is analysis of effects of diagonal disorder (random site energy) on charge density wave (CDW) and superconducting (SS) orderings as well as phase separated states in local electron pair systems. The systems considered are described by the model of hard-core charged bosons on a lattice (paulions) being equivalent to the effective pseudospin anisotropic Heisenberg model ($s = 1/2$) in the presence of uniaxial random fields and with a fixed magnetization in z direction [1, 2].

The mean-field (MFA) studies have been already performed for this problem in the particular case of half-filled band ($n = 1$) and repulsive intersite density-density interactions ($K > 0$) [2]. It was found that diagonal disorder strongly affects both CDW and SS and leads to a rich multicritical behaviour including lines of bicritical points, tricritical points and critical end points. In the simplest case of the absence of intersite interactions ($K = 0$) the question of two-dimensional superfluidity and localization in the disordered hard-core boson model has been also extensively studied numerically with the use of several complementary methods [3-8], including quantum Monte-Carlo (QMC) simulations [3-4], finite size scaling analyses [4-6] and real-space renormalization group (RG) approach [7].

In this paper we will extend the investigations of Ref. [2] and analyse the phase diagrams and ground state characteristics of the model considered for arbitrary particle concentration ($0 < n < 2$) and arbitrary (repulsive and attractive) intersite interactions ($K > 0$, $K < 0$).

The problem studied is of direct relevance for various classes of doped CDW systems with alternating valency, bipolaronic systems as well as several groups of nonconventional (exotic) superconducting materials with very short-coherence length, including the doped barium bismuthates, alkali fullerenes, Chevrel phases and underdoped cuprates. For a detailed summary of theoretical and experimental arguments in favor of the local electron pairing description of the extreme type II superconductors, we refer the reader to Refs. [1, 9-11].

The effect of randomness in a quantum system of interacting bosons is also of great interest for the subject of superfluidity of $^4$He [12, 13], superconductor-insulator transitions in thin films and vortex dynamics in type-II superconductors. Moreover, strongly disordered conventional superconductors may be also viewed as disordered (composite) boson systems [12, 14, 15]. One of the most direct experimental realization of a disordered boson system is perhaps the experiment on $^4$He in Vycor glass and in other porous media [13, 16, 17]. In these systems the porous medium provides the random potential experienced by $^4$He atoms.

The paper is organized as follows. In Sec. 2 we introduce the basic definitions and give details of the mean-field approach. We present the free energies of various possible states as well as the selfconsistent equations derived within MFA determining the order parameters and chemical potential. Sec.3 includes numerical calculations of the phase diagrams and ground state characteristics of the model in the case of a two-delta distribution of the random site energies. The phase diagrams are determined as a function of the particle concentration, the intersite density-density interaction and the strength of random on-site energies. To demonstrate the dependence of the properties of the model on the site-energy distribution function, in. Sec. 4 we repeat the calculations with the square distribution. In Sec. 5 we focus on the properties of superconducting phase and analyse the behaviour of superfluid density as well as the evolution of the superconducting critical temperature with increasing disorder. The last section is devoted to conclusion and supplementary discussion.

## 2. General formulation

The system of local electron pairs on a lattice, studied in this paper, is described by following Hamiltonian [1, 2, 18]:

$$H = -\frac{1}{2}\sum_{ij} J_{ij}\left(\rho_i^+ \rho_j^- + H.c.\right) + \sum_{ij} K_{ij}\rho_i^z \rho_j^z + \sum_i \left(E_i - \mu\right)(2\rho_i^z + 1), \quad (1)$$

where $J_{ij}$ denotes the transfer integral for the pairs from sites $\mathbf{R_i}$, $\mathbf{R_j}$ (i.e. the charge exchange interaction) $K_{ij}$ is the intersite density-density interaction, $\mu$ stands for chemical potential and $E_i$ is the random site energy. The pseudospin operators satisfy the commutation rules of the $s = 1/2$ operators

$$[\rho_i^+, \rho_j^-] = 2\rho_i^z \delta_{i,j}, \qquad [\rho_i^+, \rho_j^z] = -\rho_i^+ \delta_{i,j}, \qquad [\rho_i^-, \rho_j^z] = \rho_i^- \delta_{i,j}, \quad (2)$$

$$\rho_i^+ \rho_i^- = \rho_i^z + \frac{1}{2}, \qquad (\rho_i^+)^2 = (\rho_i^-)^2 = 0, \qquad \rho_i^+ \rho_i^- + \rho_i^- \rho_i^+ = 1. \quad (3)$$

Equation (1) has the form of an anisotropic Heisenberg model with $s=1/2$ in an effective external field $\mu$ and random field $E_i$ in the $z$ direction, that the average pseudospin magnetization has a fixed value equal to

$$\ll \frac{1}{N}\sum_i (\rho_i^z + \frac{1}{2}) \gg_{av} = \bar{n}. \quad (4)$$

$\bar{n} = n/2 = N_p/N$ is the concentration of pairs and $N$ is the total number of lattice sites and $\ll ... \gg_{av}$ means the double thermal-configuration average.

The pseudospin operators can be considered as the boson operators but with the infinite hard-core interaction. Using the representation

$$\rho_i^- = b_i, \qquad \rho_i^+ = b_i^\dagger, \qquad \rho_i^z = -\frac{1}{2} + b_i^\dagger b_i, \quad (5)$$

the Hamiltonian (1) takes the form

$$H = -\frac{1}{2}\sum_{ij} J_{ij}\left(b_i^+ b_j + H.c.\right) + \sum_{ij} K_{ij} n_i n_j + \sum_i \left(E_i - \bar{\mu}\right)n_i + const., \quad (6)$$

where $n_i = b_i^\dagger b_i$, $\bar{\mu} = 2\mu + K_o$, $const = NK_o/4$, $K_o = \sum_j K_{ij}$. The hard-core boson operators satisfy the Pauli commutation relations

$$[b_i^\dagger, b_j] = (2n_i - 1)\delta_{i,j} , \qquad (b_i^\dagger)^2 = (b_i)^2 = 0, \qquad b_i^\dagger b_i + b_i b_i^\dagger = 1 . \qquad (7)$$

The hard-core condition (7) [or (3)] allows the only single-boson occupancy of a given lattice site. The number of bosons per lattice site $\bar{n} = n/2$ is simply given by

$$<<\frac{1}{N}\sum_i n_i>>_{av} = \bar{n} \qquad (8)$$

and it determines the chemical potential.

Model (1) can be derived from the extended Hubbard model with on-site attraction and diagonal disorder in the strong-coupling limit[1, 2, 18] by the second-order perturbation theory. In this case we obtain the system of a tightly bound local pairs with the charge $\bar{e} = 2e$ and charge operators are the composite boson operators, $J_{ij} = 2t^2_{ij}/|U|$, $K_{ij} = J_{ij} + 2 W_{ij}$, where $t_{ij}$ is the hopping integral, $|U|$ - the on-site attractive interaction and $W_{ij}$ is the intersite density-density interaction between the tight binding electrons, $n = N_e/N$. The pseudospin operators $\rho_i^\alpha$, $\alpha = +, -, z$, operating in a subspace of states of double occupied and empty sites, are related to the original fermion operators through $\rho_i^+ = c_{i\uparrow}^\dagger c_{i\downarrow}^\dagger$, $\rho_i^z = \frac{1}{2}(c_{i\uparrow}^\dagger c_{i\uparrow} + c_{i\downarrow}^\dagger c_{i\downarrow} - 1)$, $\rho_i^- = (\rho_i^+)^\dagger$, where $c_{i\sigma}^\dagger$, $c_{i\sigma}$ are the electron operators.

In general, the model (1) is common to study superconductivity and CDW formation in systems with bound electron pairs, of either on-site or intersite nature, for description of the magnetic bipolarons and the case of large bipolarons in a dilute limit [1, 9, 11]. Also, the model (1), (6) is particularly useful for the extreme type II superconductors with a short coherence length, as far as the effects of phase fluctuations are concerned.

The probability distribution $P(\{E_i\})$ of $\{E_i\}$ is assumed to be

$P(\{E_i\}) = \Pi\, p(E_i)$ ,

with $p(E_i) = p(-E_i)$ . $\qquad (9)$

We will consider two types of energy distribution:

the two-delta distribution

$p(E_i) = \frac{1}{2}[\,\delta(E_i - E_o) + \delta(E_i + E_o)] \qquad (10a)$

and the rectangular distribution

$p(E_i) = 1/2E_o$ , for $|E_i| \leq E_o$,
$\quad\quad\quad\ = 0$ , otherwise. $\qquad (10b)$

In the analysis of the model we adopt the mean field variational approach (MFA) [2, 18, 19]. For a given fixed configuration of the random site energy $\{E_i\}$ the effective trial Hamiltonian is of the form:

$$H_o = -\frac{1}{2}\sum_i \Omega_i^- \left(\rho_i^+ + \text{H.c.}\right) + \sum_{ij} K_{ij}\left(\psi_i^z \rho_j^z + \rho_i^z \psi_j^z\right) + 2\sum_i (E_i - \mu)\left(\rho_i^z + \frac{1}{2}\right) +$$

$$+ \frac{1}{2}\sum_{ij}\left(J_{ij}\psi_i^+ \psi_j^- + \text{H.c.}\right) - \sum_{ij} K_{ij}\psi_i^z \psi_j^z, \qquad (11)$$

where

$$\Omega_i^\pm = \sum_j J_{ij}\psi_j^\pm, \qquad \Omega_i^z = -\sum_j K_{ij}\psi_j^z + \mu - E_i, \qquad \psi_i^\alpha = <\rho_i^\alpha>, \qquad \alpha = \pm, z. \qquad (12)$$

The thermodynamic potential is given by

$$\Omega(\{E_i\}) = -\frac{1}{\beta}\ln\left[\text{Tr}\,(e^{-\beta H_o})\right],$$

where $\beta = 1/k_B T$. From $\Omega$ the free energy per site is

$$F_o(\{E_i\}) = \frac{1}{N}\Omega(\{E_i\}) + \mu\frac{1}{N}\sum_i <2\rho_i^z + 1>_o, \qquad (13)$$

where $<...>_o = \text{Tr}\{\exp(-\beta H_o)...\}/\text{Tr}\{\exp(-\beta H_o)\}$.

After diagonalisation of $H_o$ one obtains the following form of $F_o(\{E_i\})$:

$$NF_o(\{E_i\}) = -\frac{1}{\beta}\sum_i \ln 2\cosh(\beta\Delta_i) + \frac{1}{2}\sum_{ij}\left(J_{ij}\psi_i^+\psi_j^- + \text{H.c.}\right) - \sum_{ij} K_{ij}\psi_i^z\psi_j^z + \mu\,(n-1)N + \text{const.}, \qquad (14)$$

where $\Delta_i = \left[(\Omega_i^z)^2 + (\Omega_i^+)^2\right]^{1/2}$. $\qquad (15)$

From $\partial F_o/\partial \psi_j^\alpha = 0$ one gets a set of equations determining $\psi_i^\alpha$ and $\mu$:

$$\psi_i^- = \frac{\Omega_i^-}{2\Delta_i}\text{th}(\beta\Delta_i), \qquad \psi_i^z = \frac{\Omega_i^z}{2\Delta_i}\text{th}(\beta\Delta_i), \qquad \frac{n-1}{2} = \frac{1}{N}\sum_i \psi_i^z. \qquad (16)$$

The thermodynamic potential, the free energy and the selfconsistent equations (16) should then be further configurationally averaged over the random variable $\{E_i\}$ according to a preset probability distribution $P(\{E_i\})$ (Eq.(9)) as

$$<...>_{av} = \int_{-\infty}^{\infty}\prod_i dE_i\, P(\{E_i\})\,... \,. \qquad (17)$$

Assuming the existence of two interpenetrating sublattices A and B and restricting the analysis to the two-sublatice solutions we define the order parameters as the following quenched averages:

$$X_o = \frac{2}{N} \sum_i <\psi_i^+>_{av} = <\psi_A^+ + \psi_B^+>_{av}, \qquad (18a)$$

$$X_Q = \frac{2}{N} \sum_i <\psi_i^+>_{av} \exp(i\mathbf{Q} \cdot \mathbf{R_i}) = <\psi_B^+ - \psi_A^+>_{av}, \qquad (18b)$$

$$\Delta = \frac{2}{N} \sum_i <\psi_i^z>_{av} \exp(i\mathbf{Q} \cdot \mathbf{R_i}) = <\psi_B^z - \psi_A^z>_{av}, \qquad (18c)$$

which describe the superconducting orderings ($X_o$ and $X_Q$) and CDW ordering ($\Delta$) and $\mathbf{Q}$ is half the smallest reciprocal lattice vector ($Q = \frac{\pi}{a}(1,1,1)$ for cubic lattice).

Taking into account (14), (17), (18) the quenched free energy per site is derived as:

$$<F_o>_{av} = \mu\,(n-1) + J_o \frac{(X_o^2 - X_Q^2)}{4} - \frac{K_o}{4}[(n-1)^2 - \Delta^2] +$$

$$- \frac{1}{2\beta} <\ln[4\cosh(\beta\xi_{i+})\cosh(\beta\xi_{i-})]>_{av}, \qquad (19)$$

where $K_o = \sum_j K_{ij} = zK$, $J_o = \sum_j J_{ij} = zJ$, $\xi_{i\pm}^2 = (\bar{E}_i \pm \frac{1}{2}\Delta K_o)^2 + \frac{J_o^2}{4}(X_o^2 \pm X_Q^2)$,

$$\bar{E}_i = \mu - E_i - \frac{K_o}{2}(n-1), \qquad (20)$$

z is the number of nearest neighbbours (nn),
and the equations determining order parameters and $\mu$ are:

$$\frac{\delta<F_o>_{av}}{\delta\Delta} = 0, \quad \frac{\delta<F_o>_{av}}{\delta\mu} = 0, \quad \frac{\delta<F_o>_{av}}{\delta X_q} = 0, \quad q = 0,\ Q. \qquad (21)$$

In the following we will analyse the phase diagrams and the ground state characteristics of the model (1) as a function of interaction parameters, disorder strength and particle density. In Table 1 we have summarized the types of ordered states in the system considered together with the corresponding order parameters. Apart from the homogeneous phases (SS, ηS, CDW, NO) there are also solutions for the phase separated (PS) states: PS1 (CDW-SS), PS2 (CDW-NO) and PS3 (particle droplets).

The free energies of the homogeneous phases (SS, ηS, CDW, NO) are determined from Eqs. (19)-(21), whereas the internal energy $\varepsilon$ of the PS states is calculated from the expression

$$\varepsilon_{PS} = m\varepsilon(n_+) + (1 - m)\varepsilon(n_-) , \tag{22}$$

$$mn_+ + (1 - m)n_- = n , \tag{23}$$

where $\varepsilon(n_\pm)$ is the value of the energy $\varepsilon = <<\hat{H}>>_{av}/N$ at $n = n_\pm$ corresponding to the lowest energy homogeneous solution, m is a fraction of the system with a charge density $n_+$, $(1-m)$ is a fraction with a density $n_-$ $(n_+ > n_-)$, and

$$\delta\varepsilon(n_-)/\delta n_- = (\varepsilon(n_+) - \varepsilon(n_-))/(n_+ - n_-) \tag{24}$$

(Maxwell construction).

For PS1 and PS2 $n_+ = 1$, whereas for PS3 $n_+ = 2 - n_-$. Taking into account (23), (24) Eq (22) can be written as

$$\varepsilon_{PS} = \varepsilon(n_+) - (n_+ - n) (\delta\varepsilon(n_-)/\delta n_-) . \tag{25}$$

Eqs. (19)-(21) and (22)-(25) can have several branches of solutions corresponding to various states enumerated in Table 1.

We have performed a detailed analytical and numerical analysis of the equations derived at T=0 ($\beta \to \infty$) and examined the phase diagrams of the system as a function of the interaction parameters, the particle concentration and the strength of disorder assuming two types of distribution functions (10a) and (10b). Some results concerning the finite temperature behaviour are also presented. To find the stable state and construct the phase diagrams we have compared numerically the energies $\varepsilon = <F_0(\beta \to \infty)>_{av}$ of all possible solutions choosing that which yield the lowest value of $\varepsilon$.

In the following we will assume $J > 0$. Notice, that due to a well known isomorphism between the planar ferromagnet and antiferromagnet for loose packed (alternating) lattice with nn interactions [20] the free energies are the same for the both signs of J: $<F(J)>_{av} = <F(-J)>_{av}$. Ferromagnetic XY order of pseudospins $\{\rho_i\}$ (for $J > 0$) corresponds to the SS phase ($X_0 \neq 0$), while the antiferromagnetic XY order (for $J < 0$) - to the ηS (η- pairing) phase ($X_Q \neq 0$). Thus, the phase diagrams for $J < 0$ can be directly obtained from those for $J > 0$ by the replacement $X_0 \to X_Q$ and vice-versa (cf. Table 1).

## 3. Two-delta distribution

We will normalize all the energies with respect to $J_o$ and redefine $K \equiv K_o/J_o$, $E \equiv E_o/J_o$, $\bar{\mu} \equiv \mu/J_o$, $F = \langle F_o \rangle_{av}/J_o$ and $\beta \equiv \beta J_o$. Using (10a) we obtain from Eqs. (19), (21) the free energy and selfconsistent equations as

$$F = \bar{\mu}(n-1) + \frac{X_o^2 + X_Q^2}{4} - \frac{K}{4}[(n-1)^2 - \Delta^2] +$$

$$- \frac{1}{4\beta} \ln[8 \cosh(\beta g_1) \cosh(\beta g_2) \cosh(\beta g_3) \cosh(\beta g_4)] \;, \tag{26}$$

where

$$g_1^2 = [\bar{\mu} - E - \frac{K}{2}(n-1) + \Delta\frac{K}{2}]^2 + \frac{X_o^2 + X_Q^2}{4}, \quad g_2^2 = [\bar{\mu} - E - \frac{K}{2}(n-1) - \Delta\frac{K}{2}]^2 + \frac{X_o^2 - X_Q^2}{4},$$

$$g_3^2 = [\bar{\mu} + E - \frac{K}{2}(n-1) + \Delta\frac{K}{2}]^2 + \frac{X_o^2 + X_Q^2}{4}, \quad g_4^2 = [\bar{\mu} + E - \frac{K}{2}(n-1) - \Delta\frac{K}{2}]^2 + \frac{X_o^2 - X_Q^2}{4}.$$

$$\tag{27}$$

$$8X_o = X_o \left[ \frac{\tanh(\beta g_1)}{g_1} + \frac{\tanh(\beta g_2)}{g_2} + \frac{\tanh(\beta g_3)}{g_3} + \frac{\tanh(\beta g_4)}{g_4} \right] +$$

$$+ X_Q \left[ \frac{\tanh(\beta g_1)}{g_1} - \frac{\tanh(\beta g_2)}{g_2} + \frac{\tanh(\beta g_3)}{g_3} - \frac{\tanh(\beta g_4)}{g_4} \right],$$

$$8X_Q = -X_o \left[ \frac{\tanh(\beta g_1)}{g_1} - \frac{\tanh(\beta g_2)}{g_2} + \frac{\tanh(\beta g_3)}{g_3} - \frac{\tanh(\beta g_4)}{g_4} \right] +$$

$$-X_Q \left[ \frac{\tanh(\beta g_1)}{g_1} + \frac{\tanh(\beta g_2)}{g_2} + \frac{\tanh(\beta g_3)}{g_3} + \frac{\tanh(\beta g_4)}{g_4} \right],$$

$$4\Delta = \left(\bar{\mu} - E - \frac{K}{2}(n-1)\right) \left( \frac{\tanh(\beta g_1)}{g_1} - \frac{\tanh(\beta g_2)}{g_2} \right) +$$

$$+ \left(\bar{\mu} + E - \frac{K}{2}(n-1)\right) \left( \frac{\tanh(\beta g_3)}{g_3} - \frac{\tanh(\beta g_4)}{g_4} \right) +$$

$$+ \Delta \frac{K}{2} \left( \frac{\tanh(\beta g_1)}{g_1} + \frac{\tanh(\beta g_2)}{g_2} + \frac{\tanh(\beta g_3)}{g_3} + \frac{\tanh(\beta g_4)}{g_4} \right),$$

$$4(n-1) = \left(\bar{\mu} - E - \frac{K}{2}(n-1)\right) \left( \frac{\tanh(\beta g_1)}{g_1} + \frac{\tanh(\beta g_2)}{g_2} \right) +$$

$$+ \left( \bar{\mu} + E - \frac{K}{2}(n-1) \right) \left( \frac{\tanh(\beta g_3)}{g_3} + \frac{\tanh(\beta g_4)}{g_4} \right) +$$

$$+ \Delta \frac{K}{2} \left( \frac{\tanh(\beta g_1)}{g_1} - \frac{\tanh(\beta g_2)}{g_2} + \frac{\tanh(\beta g_3)}{g_3} - \frac{\tanh(\beta g_4)}{g_4} \right). \qquad (28)$$

After taking the limit $T \to 0$ the internal energies $F(\beta \to \infty) = \varepsilon/J_o$, describing various homogeneous phases, and the energies (22)-(25) of the PS states are calculated and compared for given values of K, E and n to determine the stable phase. The resulting ground state phase diagrams as a function of interactions and particle concentration are presented in Figs. 1-4.

The phase diagram in the absence of disorder is shown in Fig. 1. For $K_o/J_o > 1$ it consists of the CDW phase for n = 1, the PS1 state for $1 > n > n_c$ and $1 < n < \bar{n}_c$ and the SS phase for $n < n_c$ and $n > \bar{n}_c$, where $n_c = \sqrt{\frac{K_o - J_o}{K_o + J_o}}$ and $\bar{n}_c = 2 - n_c$ are the critical concentrations at which the first order transitions between the PS1 state and the SS phase take place.

For $-1 < K_o/J_o < 1$ the SS phase and for $K_o/J_o < -1$ the state of particle droplets (PS3) are the ground states for all particle concentrations. Notice that for the case of nn interactions the PS1 state is strictly degenerated with the M phase in the whole range of stability of both these states [19]. As it follows from our analysis any diagonal disorder (similarly as any longer range attractive interaction $K_{ij}$ [19]) removes this degenerancy favoring the PS1 state.

The ground state diagram resulting in the presence of disorder is sketched in three dimensional projection in Fig. 2, whereas Figs. 3 and 4 present several two-dimensional plots clarifying details of the diagram.

For $-1 < K_o/J_o < 1$ (Figs. 2 and 3a) the ground state is superconducting (SS for $J_o > 0$ or $\eta S$ for $J_o < 0$) for any n if $E_o = 0$ and the increasing disorder induces a transition to nonordered (Bose glass) state. For n = 1 the transition is continuous, whereas for n ≠ 1: discontinuous (1$^{st}$ order). This is clearly seen from Fig. 5 where $X_o$ is plotted as a function of $E_o/J_o$ for a few fixed values of n.

For $K_o/J_o > 1$ (Figs. 2, 3b, c and Fig. 4) the form of the diagrams is more involved. For n = 1 (Fig. 4) the ground state is CDW if $E_o = 0$ and with increasing disorder the system exhibits either a sequence of transitions: CDW $\to$ SS $\to$ NO (for $1 < K_o/J_o < 2$) or a single $1^{st}$ order transition: CDW $\to$ NO (for $K_O/J_O > 2$). For $n \neq 1$ the ground state in the absence of disorder is either CDW–SS (PS1) (for $n_c < n < 1$) or SS (for $0 < n < n_c$), and the increasing $E_O$ can generate the following sequences of transitions: PS1 $\to$ SS $\to$ NO, PS1 $\to$ PS2 $\to$ NO, SS $\to$ PS2 $\to$ NO, or PS1 $\to$ SS $\to$ PS2 $\to$ NO (Figs. 3b, c).

For $K_O/J_O < -1$ (Fig. 3d, e and Fig. 4) at half-filling the diagram is formally identical as that for $K_O/J_O > 1$, with the replacement of CDW by PS3 (Fig. 4), while for $n \neq 1$ the ground state is PS3 if $E_O = 0$ and the increasing disorder induces either a sequence of transitions PS3 $\to$ SS $\to$ NO (it can occur only if $K_O/J_O > -2$) or a single discontinuous transition PS3 $\to$ NO (cf. Figs. 3d, e).

Finally, in Fig. 3f we have shown the ground state diagram obtained for $J_O = 0$. Since the absence of pair hopping excludes possibility of superconducting order, the diagram consists only of PS2 (CDW, if $n = 1$) and NO states, for $K_O > 0$, and PS3 and NO, for $K_O < 0$.

From the phase diagrams obtained one can draw some general conclusions concerning the effects of disorder in the system considered:

1. For any $K_O/J_O$ the minimal strength of disorder which is able to suppress superconductivity is the largest for $n = 1$ and it is strongly diminished with increasing $|n - 1|$.

2. For $n = 1$ in a definite range of repulsive interactions ($1 < K_O/J_O < 2$) the increasing disorder can destroy CDW and induce SS phase which survives randomness out to a critical amount of disorder (cf. Fig. 4). This state can be called **a disorder induced superconductivity**.

3. For attractive interactions ($K < 0$), if $-2 < K_O/J_O < -1$ the superconductivity induced by disorder, i.e. the transition from PS3 state into SS phase with increasing $E_O$, can take place also beyond half-filling ($n \neq 1$) (cf. Fig. 3d).

4. On the contrary, in the case of repulsive interactions for $n \neq 1$ the disorder reduces strongly the stability regions of the SS and PS1 states, while the nonsuperconducting CDW state (PS2) is more stable to its influence if $K_O/J_O > 1$ (cf. Fig. 3b, c). Thus, the system can exhibit phenomenon which we call **a disorder induced charge ordering** (the discontinuous transitions from the SS and PS1 states into PS2 state with increasing $E_O$).

## 4. Rectangular distribution

For the rectangular distribution function (10b) and with $J_o$ chosen as the energy unit the free energy (19) takes the form

$$F = \bar{\mu}(n-1) + \frac{X_o^2 + X_Q^2}{4} - \frac{K}{4}[(n-1)^2 - \Delta^2] - \frac{1}{4E\beta}\int_{-E}^{E} \ln[4\cosh(\beta g_1)\cosh(\beta g_2)]\,d\varepsilon, \quad (29)$$

$$g_1^2 = [\bar{\mu} - \varepsilon - \frac{K}{2}(n-1) + \Delta\frac{K}{2}]^2 + \frac{X_o^2 + X_Q^2}{4}, \quad g_2^2 = [\bar{\mu} - \varepsilon - \frac{K}{2}(n-1) - \Delta\frac{K}{2}]^2 + \frac{X_o^2 - X_Q^2}{4},$$

(30)

and the selfconsistent equations (21) are given by

$$X_o = \frac{1}{8E}\int_{-E}^{E}\left[X_o\left(\frac{\tanh(\beta g_1)}{g_1} + \frac{\tanh(\beta g_2)}{g_2}\right) + X_Q\left(\frac{\tanh(\beta g_1)}{g_1} - \frac{\tanh(\beta g_2)}{g_2}\right)\right]d\varepsilon,$$

$$X_Q = \frac{1}{8E}\int_{-E}^{E}\left[-X_o\left(\frac{\tanh(\beta g_1)}{g_1} - \frac{\tanh(\beta g_2)}{g_2}\right) - X_Q\left(\frac{\tanh(\beta g_1)}{g_1} + \frac{\tanh(\beta g_2)}{g_2}\right)\right]d\varepsilon,$$

$$4E\Delta = \int_{-E}^{E}\left(\bar{\mu} - E - \frac{K}{2}(n-1)\right)\left(\frac{\tanh(\beta g_1)}{g_1} - \frac{\tanh(\beta g_2)}{g_2}\right)d\varepsilon +$$

$$+\int_{-E}^{E}\Delta\frac{K}{2}\left(\frac{\tanh(\beta g_1)}{g_1} + \frac{\tanh(\beta g_2)}{g_2}\right)d\varepsilon,$$

$$4(n-1) = \int_{-E}^{E}\left(\bar{\mu} - \varepsilon - \frac{K}{2}(n-1)\right)\left(\frac{\tanh(\beta g_1)}{g_1} + \frac{\tanh(\beta g_2)}{g_2}\right)d\varepsilon +$$

$$+\int_{-E}^{E}\Delta\frac{K}{2}\left(\frac{\tanh(\beta g_1)}{g_1} - \frac{\tanh(\beta g_2)}{g_2}\right)d\varepsilon. \quad (31)$$

At $T = 0$ the analysis of Eqs. (29), (31) and (22), (25) analogous to that performed in previous section yields for $J_o \neq 0$ the phase diagram in three dimensional projection shown in Fig. 6 and for $J = 0$ the diagram given in Fig. 3f.

If $J = 0$ (Fig 3f, dashed lines) the SS phase is absent and disorder induces the transition: PS2 (CDW, if $n = 1$) $\to$ NO for $K > 0$, and PS3 $\to$ NO, for $K < 0$.

For $\infty > K_o/J_o > 1$ at half-filling the ground state is CDW for $E_o = 0$ and the increasing disorder provides the discontinuous (1st order) transition to the SS phase. Beyond half-filling for $K_o/J_o > 1$ the ground state is, depending on n, either PS1 or SS for $E_o = 0$ (cf. Fig. 1) and in the former case the system exhibits the 1st order transition from PS1 to SS phase with increasing $E_o$, whereas in the latter case it remains in SS state for any $E_o$ (Fig. 6).

For $-1 < K_o/J_o < 1$ the ground state is SS for arbitrary $E_o$ and n.

Finally, for $-\infty < K_o/J_o < -1$ the ground state is PS3 for arbitrary n and $E_o = 0$, and the increasing $E_o$ provides the discontinuous transition to the SS phase.

Notice that, the phase diagrams derived for rectangular distribution for $J \neq 0$ indicate a stronger suppression of CDW (existing in CDW and PS1 states) than superconductivity by the disorder effect not only for n = 1 but also beyond half-filling, in contrast to the results obtained for two-delta distribution. For rectangular distribution, the increasing disorder reduces superconducting ordering but cannot suppress it completely, although for large $E_o$ the $X_o$ becomes exponentially small: $X_o \sim \exp(-E_o/J_o)$ (see e.g. Fig. 6 in Ref. [2]).

## 5. Properties of superconducting phase

Let us consider the case $-1 < K_o/J_o < 1$. As we have seen in previous sections, only the SS order can develop in the system for this range of interactions for any n. We will focus on the behaviour of the superfluid stiffness at T = 0 and the evolution of the superconducting critical temperature $T_c$ with increasing disorder.

Within MFA the paramagnetic part of the kernel in the $\mathbf{q} \to 0$ limit vanishes at T = 0 and the London penetration depth $\lambda_L$ being directly related to the superfluid stiffness $\rho_s$ takes the form [10]:

$$\lambda_L^{-2}(0) = \frac{4\pi \bar{e}^2}{\hbar^2 c^2 a} \rho_s(0) = -K^{dia} \, (T=0), \tag{32}$$

where the diamagnetic kernel $K^{dia}$ is given by

$$K^{dia} = \frac{8\pi \bar{e}^2}{\hbar^2 c^2 a} \frac{1}{z} \frac{1}{N} \ll -\sum_k J_k \rho_k^+ \rho_k^- \gg_{av} = \frac{8\pi \bar{e}^2}{\hbar^2 c^2 a} \left( -J_o \frac{X_o^2}{z} \right), \tag{33}$$

$\bar{e} = 2e$, "a" is a lattice constant and "z" the number of n.n., $\rho_\mathbf{k}^\pm$ and $J_\mathbf{k}$ are the space-Fourier transform of $\rho_i^\pm$ and $J_{ij}$, respectively.

For the SS phase the $X_o$ and $\mu$ are is determined by a set of equations

$$X_o = \frac{1}{2} J_o < X_o \frac{\tanh \beta \xi_i}{\xi_i} >_{av}, \quad n - 1 = < \bar{E}_i \frac{\tanh \beta \xi_i}{\xi_i} >_{av}, \tag{34}$$

where $\quad \xi_i^2 = \overline{E}_i^2 + \frac{1}{4}J_o^2 X_o^2 \ , \qquad \overline{E}_i = \mu - E_i - \frac{K_o}{2}(n-1) \ ,$ (35)

and the quenched free energy of this phase is

$$<F_o>_{av} = \mu\,(n-1) + J_o\frac{X_o^2}{4} - \frac{K_o}{4}(n-1)^2 - \frac{1}{\beta}<\ln 2\cosh(\beta\xi_i)>_{av} \ . \qquad (36)$$

The results of numerical evaluation of Eqs. (32) - (36) for the random site energy distribution (10a) and (10b) are presented in Figs. (7) and (8).

In Figs 7 we show the superfluid stiffness $\rho_s(0)/2J$ (i.e. the dimensionless $\lambda_L^{-2}$) versus $|n-1|$ for several fixed values of $E_O/J_O$. The $\rho_S$ vs n plots, similarly as all the phase diagrams, are perfectly symmetric about the half-filling level, because of the particle-hole symmetry of the model [2]. Close to the half-filling point the superfluid stiffness shows a parabolic-like dependence on particle concentration for both types of disorder distribution function. For rectangular distribution (Fig. 7b) this kind of behaviour extends up to $|n-1| = 1$. On the contrary, in the case of two-delta distribution (Fig. 7a) the increasing deviation from half filling induces a discontinuous transition to the NO state for any fixed $E_O/J_O$ from the range $0 < E_O/J_O < 0.5$ (for $E_O/J_O > 0.5$ the SS phase is suppressed for any n, cf. Fig 3a).

Figs 8 show the finite temperature phase diagrams determined numerically from Eqs. (34) - (36) for several values of n. The critical temperatures of second order transition to NO state are determined from Eqs. (34) and (10) taking the limit $X_O \to 0$, whereas the first order phase boundaries (dashed lines in Fig. 8a) are obtained by comparing the free energies (36) of SS phase ($X_O \neq 0$) and NO phase ($X_O = 0$). Within MFA the SS-NO phase boundaries remain the same for any $-1 < K_O/J_O < 1$.

For rectangular distribution (Fig. 8b) the superconducting transition is of second order for any particle concentration and arbitrary $E_O/J_O$. For two-delta distribution (Fig. 8a) beyond half-filling the transition can be either second or first order, depending on the value of $E_O/J_O$. The increasing disorder changes first the nature of the phase transition from a continuous to a discontinuous type, resulting in the tricritical point (TCP), then it suppresses superconductivity for low concentrations. Finally, for large $E_O$ ($E_O/J_O > 0.5$) the system remains in a normal state at all temperatures and any n.

## 6. Summary and outlook

We have studied the effects of random site energies on superconducting and CDW orderings in local pair systems, which are described by the model of hard-core charged bosons on a lattice being equivalent to the effective pseudospin anisotropic Heisenberg model (s = 1/2) in the presence of unixial random fields and a fixed magnetization in z direction.

We have explicitly determined the complete MFA phase diagrams of this model for arbitrary particle concentration and interactions with a two-delta and a rectangular distribution of the random parameters. Depending on the strength of random on-site energies, the form of the distribution function, the intersite density-density interaction and the concentration the model can exhibit several different states including homogeneous phases: CDW, SS and Bose-glass (NO) as well as 3 types of phase separated states: CDW-SS (PS1), CDW-NO (PS2) and the state of particle droplets (PS3).

It is clearly seen that both the superconductivity and the CDW of tightly bound local electron pairs can be substantially suppressed by the disorder effects. This is in obvious contrast with the weak coupling BCS superconductors, which according to Anderson's theorem [21] are rather insensitive to diagonal disorder (nonmagnetic impurities).

At half-filling one finds a strong suppression of CDW order with random one-body potential, whereas the SS phase is relatively robust. Thus, for $K_o/J_o > 1$ increasing disorder can induce a transition from CDW to the SS phase and we call this phenomena a **disorder induced superconductivity.**

Beyond half-filling just the opposite behaviour can be observed for the two-delta distribution function: the discontinuous ($1^{st}$ order) transitions from the superconducting (SS or PS1) states into the CDW-NO state (PS2) with increasing disorder and we call it a **disorder induced charge ordering.**

Let us stress that the detailed features of the phase diagrams are rather sensitive to the choice of the distribution function. In particular for the case of rectangular distribution the model does not exhibit the disorder induced CDW and there is no tricritical point. The latter result is consistent with the Landau-type argument that TCP does not exist for any finite distribution of quenched random fields without a minimum at zero [22].

In the case of two-delta distribution there is, for any particle concentration, a critical amount of disorder, below which the superconducting order can be stable. As we have found, for any $K_o/J_o$ the critical disorder is the largest close to half-filling and strongly diminishes with increasing |n - 1|. This result can be easily explained. In fact, since the

half-filled system has the lowest kinetic energy without disorder, the critical disorder should be a maximum.

The validity of the results derived from the MFA in this paper is the same as that for spin systems in random fields. Since the low-lying transverse excitations are neglected in the MFA it seems to us that the MFA underestimates the influence of unixal field ($E_i$) on the transverse phase (SS) for $d \leq 3$ systems. Nevertheless, it is reasonable to accept that the general qualitative features calculated here are correct.

As we have pointed out in the introduction, the model (1) with $K = 0$ has been investigated numerically for $d = 2$ and $d = 3$ lattices by QMC simulations [3, 4, 6], exact diagonalizations of small systems [5] as well as with the use of quantum real-space RG method [7]. The results of these works, concerning the evolution of critical disorder with particle concentration [3, 5] and also the evolution of superfluid stiffness $\rho_s$ with n for different degrees of disorder [3], are in good qualitative agreement with our findings for this particular case (comp. e.g. Figs 7a, b with Fig. 2 in Ref[3]).

For $K_0/J_0 = 1$ the model considered describes the strong coupling limit of the attractive Hubbard model [1, 2, 18 ]. The effects of diagonal disorder in the latter model at half-filling has been recently studied with QMC techniques for $d = 2$ by Huscroft and Scalettar [23]. In agreement with our results for $K_0/J_0 = 1$ and $n = 1$ (Figs. 4 and 6) the authors find that CDW order is immediately destroyed by a random one body potential, whereas superconducting order is relatively robust and survives randomness out to a critical amount of disorder.

There are 3 types of phase separated states in the system considered. Two of them (PS1 and PS2) can appear in the case of repulsive nn interaction ($K_0/J_0 > 1$ beyond half filling), the third one (PS3) - in the case of attractive interaction ($K_0/J_0 < -1$). In the phase separated states the system breaks into coexisting domains of two different charge densities $n_+$ and $n_-$. In real systems the sizes of the domains will be finite and determined by the long-range Coulomb repulsion and structural imperfections.

We expect, on the basis of our previous studies concerning the case without disorder [19], that in the presence of long-range Coulomb interactions the general structure of the derived phase diagrams will remain unchanged except the eventual replacement of the PS1 and PS2 states by the homogeneous SS-CDW phase or the incommensurate (or striped) CDW phases.

Our results are of relevance for various groups of materials, where the existence of local pair states have been either established or suggested [1, 2, 9-11]. In particular they may provide an explanation why the doped barium bismuthates ($BaPb_{1-x}Bi_xO_3$ and $Ba_{1-x}K_xBiO_3$) [1, 24] exhibit CDW ordering in a surprisingly large range of doping concentrations before they become superconducting. Also, a disorder stabilized CDW can be a reason why some other materials with local pairs, like $Ti_{4-x}V_xO_7$ [1, 25] remain nonsuperconducting for any accessible doping level.

## 6. Acknowledgments

We would like to thank R.Micnas and T. Kostyrko for many useful discussions. This work was supported in part by the Polish Research Committee of Science, Grant No 2P03B 03717.

# Figure captions

Fig. 1.

Ground state phase diagram of the model (1) in the absence of disorder.

Fig. 2.

Ground state diagram for the two-delta distribution function and $J \neq 0$ sketched in three dimensional projection.

Fig. 3.

Two-dimensional plots of the ground state diagrams for the two-delta distribution calculated for a) $-1 < K_o/J_o < 1$, b) $K_o/J_o = 1.5$, c) $K_o/J_o = 2.5$, d) $K_o/J_o = -1.5$, e) $K_o/J_o = -2.0$, f) $J_o = 0$, $K_o \neq 0$. In Fig 2f we have also shown by dashed lines the phase boundaries PS2-NO and PS3-NO derived for the rectangular distribution of $E_o$ (see Sec. 4)

Fig. 4.

Ground state phase diagram for the two-delta distribution plotted as a function of $E_o/J_o$ and $K_o/J_o$ for $n = 1$. Solid and dashed lines denote first and second-order transitions, respectively.

Fig. 5.

A square of superconducting order parameter $X_o^2$ at $T = 0$ as a function of increasing disorder $E_o/J_o$ plotted for the two-delta distribution, $-1 < K_o/J_o < 1$ and several fixed values of n (numbers to the curves).

Fig. 6.

Ground state diagram for the square distribution and $J \neq 0$ in three dimensional projection.

Fig. 7.

Superfluid stiffness $\frac{\rho_s}{2J} = \lambda_L^{-2} / \frac{8\pi \overline{e}^2 J}{\hbar^2 c^2 a}$ at T = 0 as a function of particle concentration n, plotted for $-1 \leq K_O/J_O \leq 1$ and several fixed values of $E_O/J_O$: a) two-delta distribution, b) rectangular distribution.

Fig. 8.

Finite temperature phase diagram as a function of $E_O/J_O$, plotted for $-1 \leq K_O/J_O \leq 1$ and several fixed values of n: a) two-delta distribution; solid and dashed lines mark second- and first-order phase transitions, respectively, open circle denotes the tricritical point (TCP). b) rectangular distribution.

Table 1. The types of ordered states

| TYPE OF ORDERED STATE | ORDER PARAMETERS |
|---|---|
| 1) SUPERCONDUCTING PHASE ($J > 0$) <br> **(SS)** | $X_O \neq 0$ |
| 2) $\eta$-PAIRING PHASE ($J < 0$) <br> **($\eta$S)** | $X_Q \neq 0$ |
| 3) CHARGE–ORDERED PHASE <br> **(CDW)** | $\Delta \neq 0$ |
| 4) PHASE SEPARATED STATE <br> of CDW and SS (or $\eta$S) <br> **(PS1)** | domains of $X_O \neq 0$ (or $X_Q \neq 0$) ($n_-$) <br> and $\Delta \neq 0$ ($n_+$) <br> ($n_+ = 1$, $n_- < 1$) |
| 5) MIXED PHASE <br> of CDW, SS and $\eta$S <br> **(M)** | $X_O \neq 0$, $X_Q \neq 0$ <br> and $\Delta \neq 0$ |
| 6) PHASE SEPARATED STATE <br> of CDW and NO <br> **(PS2)** | domains of $\Delta \neq 0$ ($n_+$) <br> ($n_+ = 1$, $n_- < 1$) |
| 7) PHASE SEPARATED STATE <br> of particle droplets <br> **(PS3)** | $n_+ - n_- \neq 0$ <br> $X_O = X_Q = \Delta = 0$ <br> ($n_+ > 1$, $n_- < 1$) |
| 8) NONORDERED STATE (BOSE GLASS) <br> **(NO)** | — |

Fig. 1

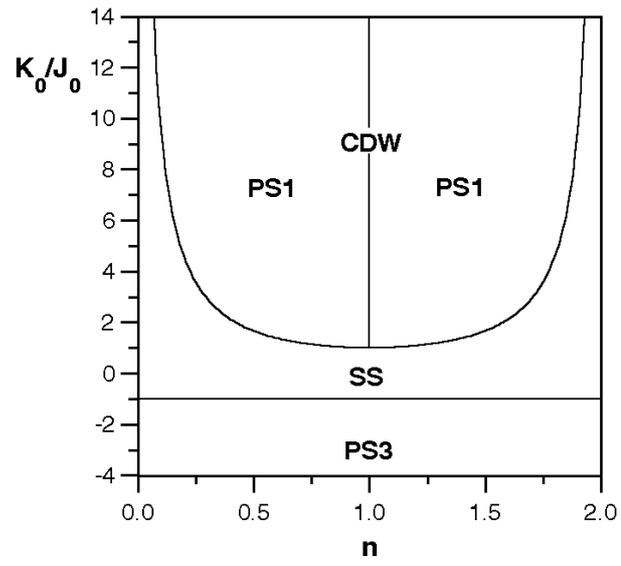

Fig. 2

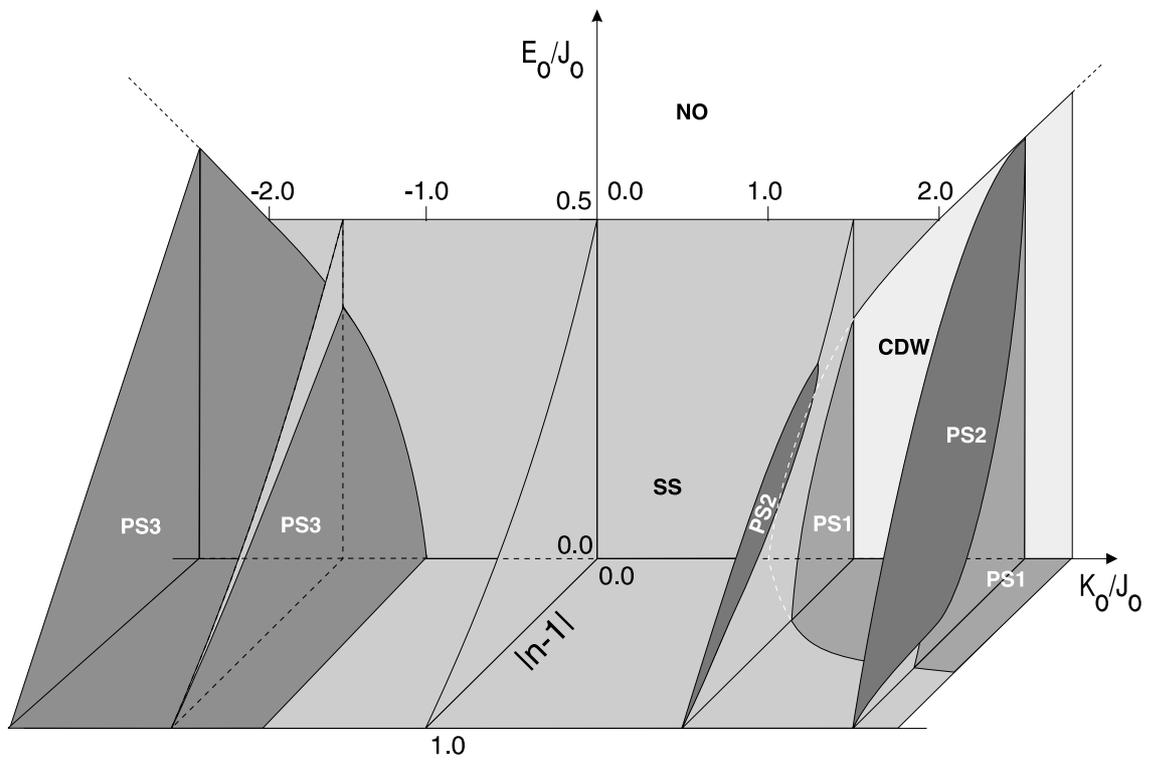

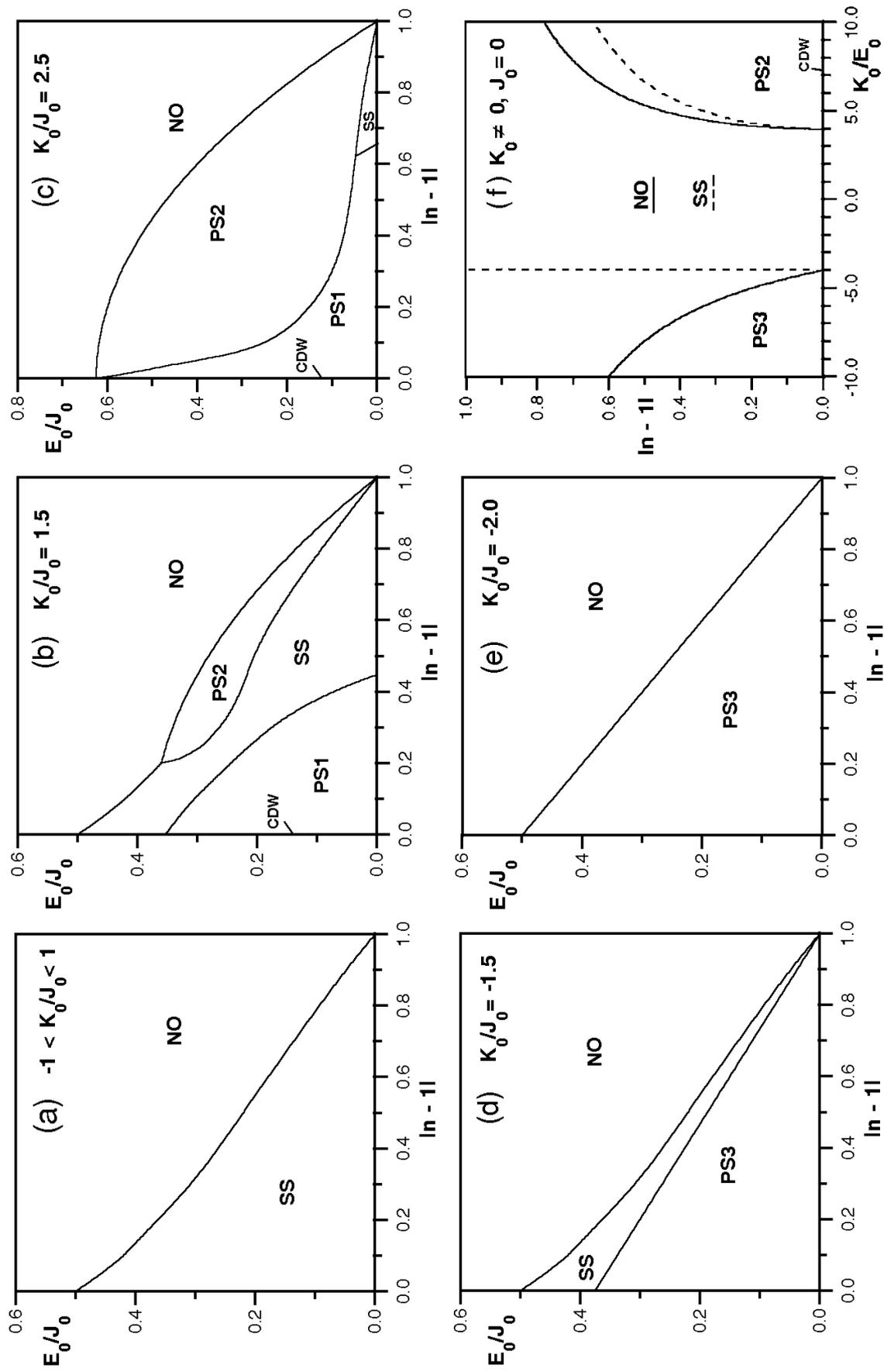

Fig. 3

Fig. 4

Fig. 5

Fig. 6

Fig. 7

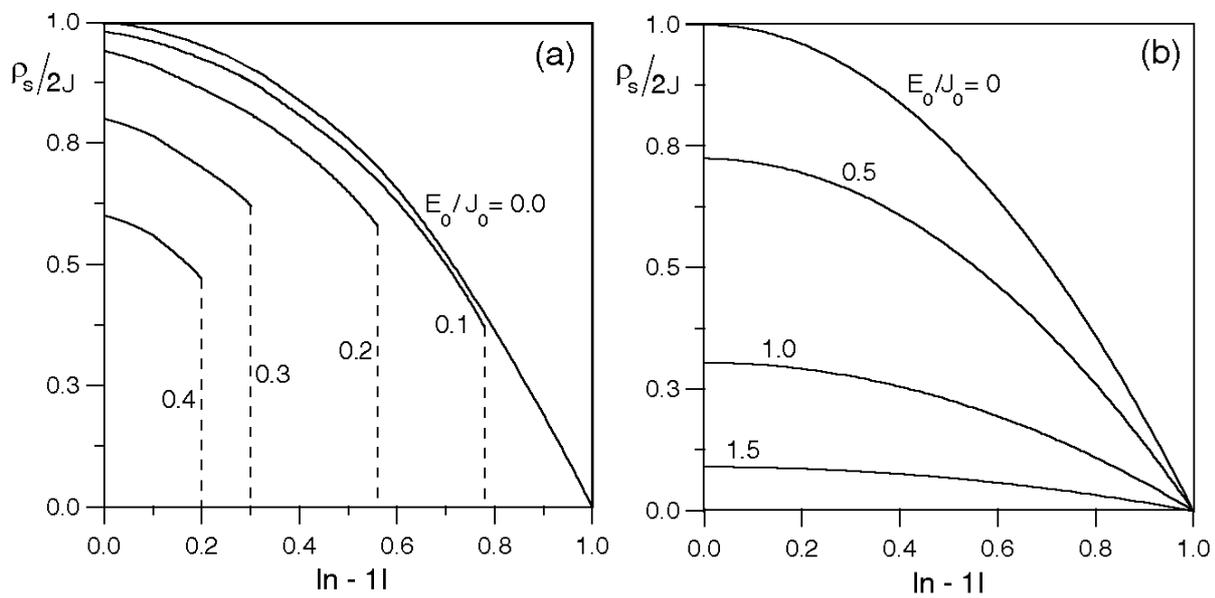

Fig. 8

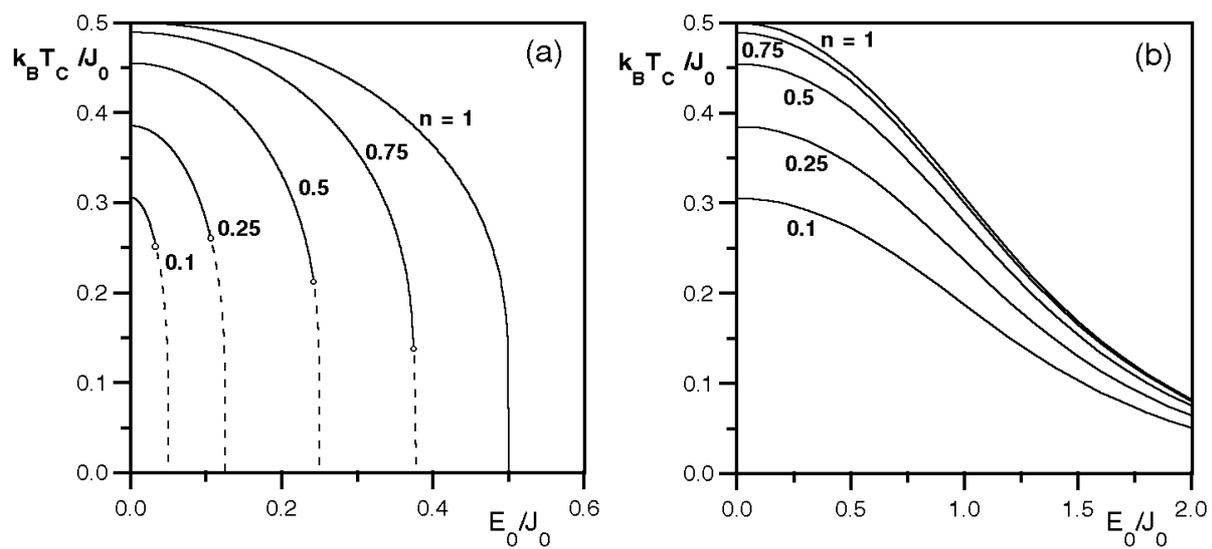